\documentclass[11pt]{article}
\usepackage{amsmath}
\usepackage{amssymb}
\usepackage{graphicx}
\usepackage{caption2}
\usepackage{amsfonts}
\usepackage{cite}

\newcommand{\be}{\begin{equation}}
\newcommand{\ee}{\end{equation}}
\newcommand{\bea}{\begin{eqnarray}}
\newcommand{\eea}{\end{eqnarray}}
\newcommand{\bef}{\begin{figure}}
\newcommand{\ef}{\end{figure}}
\newcommand{\bt}{\begin{tabular}}
\newcommand{\et}{\end{tabular}}
\newcommand{\bno}{\begin{enumerate}}
\newcommand{\eno}{\end{enumerate}}

\def\3{\ss}

\catcode`\"=\active
\def"{\accent'177}

\begin{document}

\begin{center}

{\Large\bf Reply to the comments of McMullen et al. (arXiv:2510.04828)}

\vspace{0.3cm}

Shijun Liao$^{1,2 *}$ and Shijie Qin$^{1}$

$^{1}$State Key Laboratory of Ocean Engineering, Shanghai 200240, China\\

$^{2}$School of Ocean and Civil Engineering, Shanghai Jiao Tong University, Shanghai 200240, China\\

\vspace{0.3cm}

$^*$Corresponding author: sjliao@sjtu.edu.cn

\end{center}

\hspace{-0.5cm}{\bf Abstract} { 
McMullen et al.~\cite{mcmullen2025arXiv} comment that  the numerical simulations that {\em explicitly} include random velocity fluctuations ``should exhibit a thermal-fluctuation-dominated range''  consistent with the literature, so that our results  (J. Fluid Mech. 1008, R2, 2025)~\cite{Liao-2025-JFM-PS}  ``contradict other results in the literature''.
 First of all, we would give an opposite example against this viewpoint:  DNS results (that are badly polluted by numerical noises quickly, as mention in \S~2)  {\em implicitly} include random numerical noises, but they also {\em do not} exhibit a thermal-fluctuation-dominated range.  In other words, DNS results in the literature qualitatively agree with ours at this point. 
In addition, we highly suggest that  influences of numerical noises on  statistics of turbulent flows given by {\em all} numerical approaches should be carefully checked, since numerical noises might  have huge influences on statistics of chaotic systems (including turbulence), as pointed by Lorenz~\cite{Lorenz2006Tellus} in 2006.  Detailed replies are given below.      
}

\hspace{-0.5cm}{\bf Keywords} turbulence, chaos, clean numerical simulation

\section{Introduction}

First of all, we thank  all authors (arXiv:2510.04828)~\cite{mcmullen2025arXiv} for their  comments on our paper (J. Fluid Mech. 1008, R2, 2025) \cite{Liao-2025-JFM-PS}, since deep discussions and frank exchanges of different viewpoints are helpful for the development of fluid mechanics.    

 Using a simple model (based on the deterministic Navier-Stokes equations) and clean numerical simulation (CNS) in which artificial numerical noise is negligible over a finite, sufficiently long interval of time, we show evidence  that artificial numerical noise in direct numerical simulation (DNS) of Navier-Stokes (NS) turbulence is ``approximately equivalent to  thermal fluctuation and/or stochastic environmental noise''~\cite{Liao-2025-JFM-PS}. Currently, McMullen et al.~\cite{mcmullen2025arXiv} comment that  the numerical simulations that {\em explicitly} include random velocity fluctuations \cite{Liao-2025-JFM-PS} ``should exhibit a thermal-fluctuation-dominated range''  consistent with the literature, so that our results \cite{Liao-2025-JFM-PS}  ``contradict other results in the literature''.
 
First of all, we would give an opposite example against this viewpoint:  DNS results (that are badly polluted by numerical noises quickly, as mention in \S~2)  {\em implicitly} include random numerical noises, but they also {\em do not} exhibit a thermal-fluctuation-dominated range.  In other words, DNS results in the literature qualitatively agree with ours at this point.      
 
 Our replies are briefly given below .    

\section{Reliability of numerical simulation of  turbulence}

In 1890s Poincar\'{e} \cite{poincare1890probleme} discovered that chaotic system has sensitivity dependence on initial condition, which was named  ``butterfly-effect'' by Lorenz \cite{Lorenz1963} in 1963.  More importantly,  Lorenz~\cite{Lorenz2006Tellus}  discovered in 2006  that, 
for a chaotic system,  numerical noises have large influences not only on its trajectory but also on its statistics and characteristics:  its maximum Lyapunov exponent  invariably alternates  between negative and positive values even when the time-step of traditional numerical algorithms becomes rather small. Thus, the basic characteristic of the numerical simulations alternates between chaotic and non-chaotic states, which are fundamentally quite different from each other.   A few groups \cite{Teixeira2007, Yao2008, Yao2010, Hoover2015} confirmed Lorenz's conclusions and  some  even  pointed out  that,  ``for chaotic systems, numerical convergence {\em cannot} be guaranteed {\em forever}''~\cite{Teixeira2007}, and that ``{\em all} chaotic responses are simply numerical {\em noises} and have {\em nothing} to do with differential equations'' \cite{Yao2008, Yao2010}.   These are very negative viewpoints about the reliability of numerical simulations of chaotic systems.   They  \cite{Teixeira2007, Yao2008, Yao2010, Hoover2015} illustrated that the {\em exact} solution of a chaotic system is one thing, but unfortunately its {\em numerical} simulation might be the other completely different thing, because unavoidable numerical noises  increase inexorably  to  a  macro-level due to the butterfly-effect of chaos.  

Many researchers reported that Navier-Stokes (NS) turbulence (i.e. turbulence governed by NS equations) are chaotic \cite{Deissler1986PoF, aurell1996growth, boffetta2017chaos, berera2018chaotic, boffetta2001predictability, Vassilicos2023JFM}.  So, according to the conclusions given by Lorenz~\cite{Lorenz2006Tellus} and other researchers~\cite{Teixeira2007, Yao2008, Yao2010, Hoover2015}, the reliability of numerical simulations (including DNS) of NS turbulence  (as a chaotic system) is rather suspect.  Thus, we had to answer the following fundamental questions:
\begin{enumerate}
\item[(1)] Do numerical noises have huge influences on {\em spatio-temporal  trajectory} of  numerical simulations of NS turbulence?
\item[(2)]  Are {\em statistics} of numerical simulations of NS turbulence sensitive to  numerical noises?
\end{enumerate}

To gain {\em reliable} numerical simulation of a chaotic system, Liao~\cite{Liao2009} proposed the so-called ``clean numerical simulation'' (CNS) \cite{Liao2013, Liao2014SCPMA, Hu2020JCP, Qin2020CSF, Zhang2025CPC, Liao2023book}.   The computational efficiency of CNS has been increased several orders of magnitudes so that some simple NS turbulences can be solved by means of CNS.  Briefly speaking, different from traditional numerical algorithms, CNS decreases both of truncation error and round-off error of numerical simulation to such a low level that the  numerical noise is much smaller than its ``true'' solution and thus is negligible in a time-interval $[0,T_{c}]$ that is long enough for calculating statistics, where $T_{c}$ is called ``critical predictable time''.  Note that $T_{c}$ is a key concept in CNS:  results given by CNS are  reliable within  $t \leq T_{c}$,  but otherwise not.  This is essentially different from DNS whose simulation can be {\em arbitrarily} long if one has enough resources of computation.  In fact, one can regard DNS as a special case of CNS, but unfortunately the corresponding $T_{c}$ of DNS results is too short to calculate statistics.   For more details,  please see Liao's book \cite{Liao2023book}.

Using CNS, we can do `clean' numerical experiment of NS turbulence.   Comparing CNS results with those given by DNS of the same NS turbulence, one can investigate the influence of numerical noises.  It is found  \cite{Qin2022JFM, Qin2024JOES, qin2025ultrachao-arXiv} that         
\begin{enumerate}
\item[(A)]  In all cases, DNS spatio-temporal trajectories of NS turbulence are quickly polluted by numerical noises badly, i.e.  numerical noise is mostly at the same order of magnitude as its true solution;    
\item[(B)]  In most cases,  the statistic results of DNS are the same as those given by CNS, indicating the ``statistic stability'' of these NS turbulences.  However, in some cases, numerical noises can  lead to large distinctions even in flow type and statistics, indicating that numerical noise sometimes might have large influences in statistics of NS turbulence.      
\end{enumerate}     
In addition, using CNS to solve a kind of 2D turbulent Kolmogorov flow subject to a specially chosen initial condition that contains micro-level disturbances at different orders of magnitude, Liao and Qin \cite{Liao-2025-JFM-NEC} revealed an interesting phenomenon of NS turbulence,  called ``the noise-expansion cascade'':  all micro-level disturbances at different orders of magnitude evolute and grow continuously, step by step, as an inverse cascade, to reach the macro-level, and  each disturbance could greatly change the characteristics of the 2D turbulent Kolmogorov flow\footnote[2]{The related code of CNS of the NS turbulence and some movies  can be downloaded via GitHub
({https://github.com/sjtu-liao/2D-Kolmogorov-turbulence}).}.  This highly suggests that  each disturbance must be considered in the NS turbulence, even if the disturbance is many orders of magnitude smaller than others.  Unfortunately, NS turbulence  neglects all stochastic disturbances when $t>0$.  This leads to a logic paradox in theory.        

Our above-mentioned investigations \cite{Qin2022JFM, Qin2024JOES, Liao-2025-JFM-NEC, qin2025ultrachao-arXiv} highly suggest that the following three things, i.e.  
 \begin{enumerate}
\item[(a)]  exact (or clean) solution of the NS turbulence, 
\item[(b)]  numerical simulations (that are quickly  polluted by numerical noise badly) of the NS turbulence,
\item[(c)] real turbulence in practice,
 \end{enumerate}
 might be {\em completely  different}.  A lot of theoretical, numerical and experimental  investigations are needed to  reveal  the relationships  and  differences  between  them.  
  We believe that the same conclusions should hold when  the NS turbulence is replaced by other turbulence models, such as fluctuating hydrodynamics (FHD)~\cite{Bell2022JFM},  direct simulation Monte Carlo (DSMC)~\cite{McMullen2022PRL},  molecular dynamics (MD)~\cite{Komatsu2014IJMP}, and so on, since turbulence should be  chaotic in essence and numerical noises are unavoidable for all numerical approaches.  Like DNS, these numerical approaches use double precision and/or low order algorithms so that numerical noise should quickly increase to the same order of magnitude as  true solution.    Thus, based on our experience \cite{Qin2022JFM, Qin2024JOES, qin2025ultrachao-arXiv,  Liao-2025-JFM-NEC} on NS turbulence, it is {\em questionable} whether or not  these numerical simulations agree with  `true' solution of the corresponding mathematical model in statistics.  
So, we highly suggest that the influences of numerical noises on statistic results given by {\em all} numerical approaches (including FHD, DSMC,  MD and so on) should be checked  {\em very carefully}, before they can be used as ``benchmark solution''.
  

McMullen et al.~\cite{mcmullen2025arXiv} comment that  the numerical simulations that {\em explicitly} include random velocity fluctuations ``should exhibit a thermal-fluctuation-dominated range''  consistent with the literature, so that our results~\cite{Liao-2025-JFM-PS}  ``contradict other results in the literature''.
Here, we would give an opposite example against this viewpoint:  DNS results (that are badly polluted by numerical noises quickly, as mention above)  {\em implicitly} include random numerical noises, but they also {\em do not} exhibit a thermal-fluctuation-dominated range, as mentioned in the literature.   This qualitatively agrees with our results at this point.  
  
Indeed,  our results ``contradict other results in the literature''.  We would like to emphasize here that our results are `clean' and thus reliable,  but the  influences of numerical noises on the results in the literature have not been checked very carefully.   Certainly, more investigations are necessary in future so as to give a sound conclusion.

 \section{Model for environmental noise and thermal fluctuation }
 
 The motivation of our paper  \cite{Liao-2025-JFM-PS} is to find some  {\em relationships} between {\em artificial} numerical noise and {\em physical} disturbances such as environmental noise and/or thermal fluctuation.

Until now, CNS has been successfully applied {\em only} to deterministic equations.  Thus,   in order to gain a `clean' simulation with negligible numerical noises, the deterministic NS turbulence is used in \cite{Liao-2025-JFM-PS}, since it as one millennium problem \cite{MillenniumProblem} is widely used by turbulence community.  Note that the NS turbulence itself does not include the stochastic terms about thermal fluctuation and/or environmental  noise that  can ``enter via a random stress tensor that is constrained to satisfy the fluctuation-dissipation relation''.  So, we propose in \cite{Liao-2025-JFM-PS} such a {\em new} model for influence of thermal fluctuation  and/or  environmental  noise: the simulation at  $ t_{n+1} = (n+1)\Delta t$ is gained by CNS (with negligible numerical noises) using the NS equations, which is then modified at $t_{n+1}$ by {\em suddenly} adding a stochastic velocity field, where the random field is taken to be Gaussian white noise with zero mean and standard deviation $\sigma$ so as to include thermal fluctuations and/or environmental noises.  Since the evolution from $t_{n}$ to $t_{n+1}$  is governed by the deterministic NS turbulence and the stochastic velocity field is {\em suddenly} added at $t_{n+1}$, this model is essentially deterministic.  Therefore, the numerical algorithms for stochastic differential equations are {\em unnecessary} for our model adapted in \cite{Liao-2025-JFM-PS}.   

It is true that the thermal noise strength used in \cite{Liao-2025-JFM-PS} is ``many orders of magnitude smaller than what is representative of any physically realizable turbulent flow''.  But this is not important and does not change our conclusions reported in \cite{Liao-2025-JFM-PS}, since, due to the so-called ``noise-expansion cascade''~\cite{Liao-2025-JFM-NEC}, the  thermal noise strength quickly increases to an order of magnitude having physical meaning.

Note that NS turbulence neglects all physical stochastic disturbances when $t>0$.  So, it is not an ideal model for thermal fluctuation.  Hopefully, one can solve Landau-Lifshitz-Navier-Stokes (LLNS) equations \cite{LLNS1959} (which consider the influence of thermal fluctuation) by means of CNS in the near future.  This is the reason  why  in \cite{Liao-2025-JFM-PS} we made such a conclusion that numerical noise might be {\em approximately}  equivalent to thermal fluctuation or/and  environmental noise.

\section{Concluding remarks}   

Our replies are briefly  given  below:
\begin{enumerate}
\item[(I)]  McMullen et al.~\cite{mcmullen2025arXiv}  mentioned  that  our numerical simulations \cite{Liao-2025-JFM-PS} that {\em explicitly} include random velocity fluctuations  ``should exhibit a thermal-fluctuation-dominated range''  consistent with the literature.  However,  as mention in \S~2, DNS results {\em implicitly} include random numerical noises, but they also {\em do not} exhibit a thermal-fluctuation-dominated range.  In other words, DNS results in the literature qualitatively {\em agrees}  with ours in~\cite{Liao-2025-JFM-PS} at this point.   
\item[(II)] In addition, statistic results in the literature are based on numerical simulations of chaotic systems,  which, similar to DNS results,  might be badly polluted by numerical noises, too.  So, according to our experiences on NS turbulence~\cite{Qin2022JFM, Qin2024JOES, qin2025ultrachao-arXiv, Liao-2025-JFM-NEC},  we highly suggest that the influences of numerical noises on statistic results given by {\em all} approached (including FHD, DSMC, MD and so on) should be checked  {\em very carefully}, before they can be used as `benchmark solution'.  
\item[(III)]  We propose in \cite{Liao-2025-JFM-PS} a  {\em new} simple model for influence of thermal fluctuation and/or environmental noise.    This model is {\em deterministic} in essence, and thus is {\em unnecessary} to use numerical techniques for stochastic differential equations.   Hopefully, LLNS equations can be solved by CNS in future, since it includes thermal fluctuation.          
\end{enumerate}
       
According to our experiences on NS turbulence~\cite{Qin2022JFM, Qin2024JOES, qin2025ultrachao-arXiv, Liao-2025-JFM-NEC},  we might artificially add a non-negligible influence on turbulent flows when we investigate them by numerical and/or experimental approaches, because turbulent flows are essentially chaotic.  Thus, it is very important for us to  know the conditions under which these tiny stochastic disturbances are negligible or non-negligible, the corresponding turbulent flows have statistic stability (or instability), and so on.    Hopefully, the above discussions could attract more attentions of turbulence community to investigate {\em the relationships and differences between the mathematical turbulence models, their numerical simulations, and real turbulent flows in practice}~\cite{qin2025ultrachao-arXiv}.  
 
\vspace{0.5cm}

\hspace{-0.5cm}{\bf Acknowledgements}
Thanks a lot to McMullen {\em et al.} for their comments.  Especially, Shijun would  like to express his sincere gratitude to Prof. Yao for introducing  the  interesting comments~\cite{Yao2008} about Lorenz's pioneering work~\cite{Lorenz2006Tellus} at an international conference in 2008, which gave him a shock that attracted his attention to a completely new field, i.e. the reliability of numerical simulations of chaotic systems and turbulent flows.

\bibliographystyle{elsarticle-num}
\bibliography{Kolmogorov}

\begin{thebibliography}{10}
\expandafter\ifx\csname url\endcsname\relax
  \def\url#1{\texttt{#1}}\fi
\expandafter\ifx\csname urlprefix\endcsname\relax\def\urlprefix{URL }\fi
\expandafter\ifx\csname href\endcsname\relax
  \def\href#1#2{#2} \def\path#1{#1}\fi

\bibitem{mcmullen2025arXiv}
R.~M. McMullen, M.~A. Gallis, I.~Srivastava, A.~J. Nonaka, J.~B. Bell, A.~L.
  Garcia, \href{https://arxiv.org/abs/2510.04828}{Comment on "physical
  significance of artificial numerical noise in direct numerical simulation of
  turbulence"} (2025).
\newblock \href {http://arxiv.org/abs/2510.04828} {\path{arXiv:2510.04828}}.
\newline\urlprefix\url{https://arxiv.org/abs/2510.04828}

\bibitem{Liao-2025-JFM-PS}
S.~Liao, S.~Qin, Physical significance of artificial numerical noise in direct
  numerical simulation of turbulence, J. Fluid Mech. 1008 (2025) R2.

\bibitem{Lorenz2006Tellus}
E.~N. Lorenz, Computational periodicity as observed in a simple system, Tellus
  A: Dynamic Meteorology and Oceanography 58A (2006) 549 -- 557.

\bibitem{poincare1890probleme}
H.~Poincar{\'e}, Sur le probl{\`e}me des trois corps et les {\'e}quations de la
  dynamique, Acta Math. 13~(1) (1890) A3--A270.

\bibitem{Lorenz1963}
E.~N. Lorenz, Deterministic nonperiodic flow, J. Atmos. Sci. 20~(2) (1963)
  130--141.

\bibitem{Teixeira2007}
J.~Teixeira, C.~Reynolds, K.~Judd, Time step sensitivity of nonlinear
  atmospheric models: Numerical convergence, truncation error growth, and
  ensemble design, J. Atmos. Sci. 64 (2007) 175 -- 188.
\newblock \href {http://dx.doi.org/10.1175/JAS3824.1}
  {\path{doi:10.1175/JAS3824.1}}.

\bibitem{Yao2008}
L.~Yao, D.~Hughes, Comment on ``computational periodicity as observed in a
  simple system'' by {Edward N. Lorenz} (2006), Tellus-A 60 (2008) 803 -- 805.
\newblock \href {http://dx.doi.org/10.1111/j.1600-0870.2008.00301.x}
  {\path{doi:10.1111/j.1600-0870.2008.00301.x}}.

\bibitem{Yao2010}
L.~Yao, Computed chaos or numerical errors, Nonlinear Analysis: Modelling and
  Control 15 (2010) 109--126.
\newblock \href {http://dx.doi.org/10.15388/NA.2010.15.1.14368}
  {\path{doi:10.15388/NA.2010.15.1.14368}}.

\bibitem{Hoover2015}
W.~Hoover, C.~Hoover, Comparison of very smooth cell-model trajectories using
  five symplectic and two {Runge-Kutta} integrators, Computational Methods in
  Science and Technology 21 (2015) 109 -- 116.
\newblock \href {http://dx.doi.org/10.12921/cmst.2015.21.03.001}
  {\path{doi:10.12921/cmst.2015.21.03.001}}.

\bibitem{Deissler1986PoF}
R.~G. Deissler, Is {N}avier-{S}tokes turbulence chaotic?, Phys. Fluids 29
  (1986) 1453 -- 1457.

\bibitem{aurell1996growth}
E.~Aurell, G.~Boffetta, A.~Crisanti, G.~Paladin, A.~Vulpiani, Growth of
  noninfinitesimal perturbations in turbulence, Phys. Rev. Lett. 77~(7) (1996)
  1262.

\bibitem{boffetta2017chaos}
G.~Boffetta, S.~Musacchio, Chaos and predictability of homogeneous-isotropic
  turbulence, Phys. Rev. Lett. 119~(5) (2017) 054102.

\bibitem{berera2018chaotic}
A.~Berera, R.~D. J.~G. Ho, Chaotic properties of a turbulent isotropic fluid,
  Phys. Rev. Lett. 120~(2) (2018) 024101.

\bibitem{boffetta2001predictability}
G.~Boffetta, S.~Musacchio, Predictability of the inverse energy cascade in {2D}
  turbulence, Phys. Fluids 13~(4) (2001) 1060--1062.

\bibitem{Vassilicos2023JFM}
J.~Ge, J.~Rolland, J.~C. Vassilicos, The production of uncertainty in
  three-dimensional {N}avier-{S}tokes turbulence, J. Fluid Mech. 977 (2023)
  A17.

\bibitem{Liao2009}
S.~Liao, On the reliability of computed chaotic solutions of non-linear
  differential equations, Tellus Ser. A-Dyn. Meteorol. Oceanol. 61~(4) (2009)
  550--564.

\bibitem{Liao2013}
S.~Liao, On the numerical simulation of propagation of micro-level inherent
  uncertainty for chaotic dynamic systems, Chaos Solitons Fractals 47 (2013)
  1--12.

\bibitem{Liao2014SCPMA}
S.~Liao, P.~Wang, On the mathematically reliable long-term simulation of
  chaotic solutions of {Lorenz} equation in the interval [0, 10000], Sci. China
  - Phys. Mech. Astron. 57~(2) (2014) 330 -- 335.

\bibitem{Hu2020JCP}
T.~Hu, S.~Liao, On the risks of using double precision in numerical simulations
  of spatio-temporal chaos, J. Comput. Phys. 418 (2020) 109629.

\bibitem{Qin2020CSF}
S.~Qin, S.~Liao, Influence of numerical noises on computer-generated simulation
  of spatio-temporal chaos, Chaos Solitons Fractals 136 (2020) 109790.

\bibitem{Zhang2025CPC}
B.~Zhang, S.~Liao, An automated parallel program of {Clean Numerical
  Simulation} for chaotic systems governed by {ODE}s, Computer Physics
  Communications 317 (2025) 109855.

\bibitem{Liao2023book}
S.~Liao, Clean Numerical Simulation, Chapman and Hall/CRC, 2023.

\bibitem{Qin2022JFM}
S.~Qin, S.~Liao, Large-scale influence of numerical noises as artificial
  stochastic disturbances on a sustained turbulence, J. Fluid Mech. 948 (2022)
  A7.

\bibitem{Qin2024JOES}
S.~Qin, Y.~Yang, Y.~Huang, X.~Mei, L.~Wang, S.~Liao, Is a direct numerical
  simulation {(DNS)} of {Navier-Stokes} equations with small enough grid
  spacing and time-step definitely reliable/correct?, Journal of Ocean
  Engineering and Science 9 (2024) 293 -- 310.

\bibitem{qin2025ultrachao-arXiv}
S.~Qin, K.~Xu, S.~Liao, \href{https://arxiv.org/abs/2510.06620}{Ultra-chaotic
  property of {Navier-Stokes} turbulence} (2025).
\newblock \href {http://arxiv.org/abs/2510.06620} {\path{arXiv:2510.06620}}.
\newline\urlprefix\url{https://arxiv.org/abs/2510.06620}

\bibitem{Liao-2025-JFM-NEC}
S.~Liao, S.~Qin, Noise-expansion cascade: an origin of randomness of
  turbulence, J. Fluid Mech. 1009 (2025) A2.

\bibitem{Bell2022JFM}
J.~B. Bell, A.~Nonaka, A.~L. Garcia, G.~Eyink, Thermal fluctuations in the
  dissipation range of homogeneous isotropic turbulence, J. Fluid Mech. 939
  (2022) A12.

\bibitem{McMullen2022PRL}
R.~M. McMullen, M.~C. Krygier, J.~R. Torczynski, M.~A. Gallis, {Navier-Stokes}
  equations do not describe the smallest scales of turbulence in gases, Phys.
  Rev. Lett. 128 (2022) 114501.

\bibitem{Komatsu2014IJMP}
T.~S. Komatsu, S.~Matsumoto, T.~Shimada, N.~Ito, A glimpse of fluid turbulence
  from the molecular scale, Int. J. Mod. Phys. C 25 (2014) 1450034.

\bibitem{MillenniumProblem}
{Clay Mathematics Institute of Cambridge, Massachusetts}, {The Millennium Prize
  Problems}, \url{https://www.claymath.org/millennium-problems/} (2000).

\bibitem{LLNS1959}
L.~D. Landau, E.~M. Lifshitz, Course of Theoretical Physics: Fluid Mechanics
  (Vol. 6), Addision-Wesley, Reading, 1959.

\end{thebibliography}

\end{document}